\documentclass[a4paper]{aa}

\usepackage{multirow}
\usepackage{graphicx}
\usepackage{txfonts}
\usepackage{lipsum}
\usepackage{subcaption}        
\usepackage{lscape}           
\usepackage{placeins}          
\usepackage{hyperref}
\hypersetup{colorlinks=true,linkcolor=violet,urlcolor=cyan,citecolor=blue}

\usepackage{xcolor} 
\definecolor{blueish}{HTML}{3A5C94}
\definecolor{redish}{HTML}{994040}

\begin{document}

   \title{The impact of MUSE sky subtraction on unresolved stellar population end products}
   
   \subtitle{Study cases of NGC\,1097 and NGC\,1433}

   \author{Camila de Sá-Freitas\inst{1}\corrauth{camila.desafreitas@eso.org}       
        \and Fuyan Bian\inst{1} \and Patrícia Sánchez-Blázquez\inst{2}
        }

   \institute{European Southern Observatory,  Alonso de Córdova 3107, Vitacura, Región Metropolitana, Chile
   \and Departamento de Física Teórica, Universidad Autónoma de Madrid, E-28049 Cantoblanco, Spain}

   \date{Received \today}

\abstract{One of the main steps in data reduction for ground-based astronomy is sky subtraction. For extended objects, it is necessary to observe dedicated sky frames, usually with airmass values similar to the target's, using different pointing strategies. However, it is not well documented how different observing strategies can affect the science end product. In particular, with the use of the Multi Unit Spectroscopic Explorer (MUSE/VLT), this can yield distinct stellar population products, such as age, metallicity, and $\alpha$-enhancement. To address {this}, we considered two {case studies}, NGC\,1097 and NGC\,1433, seeking to investigate: (\textit{i}) the sky stability in observations taken consecutively {over} a period of 2 hours with thin sky transparency and airmass values of 1.25-1.05, and (\textit{ii}) the impact of different sky conditions on the same target. Among our main results, we find that for galaxies with 0.0014 $\leq$ z $\leq$ 0.0077, the sky line [NI] $\lambda5200$ coincides with the Mg$_b$ bandpass, which is essential for metallicity measurements, highlighting the need for proper sky subtraction. In addition, for {regions with relative fluxes above 10 with} respect to the sky in NGC\,1097, sky frames with variation up to $\sim18\%$ still retrieve consistent stellar population properties -- in this case, this {corresponds to flux} values of $\geq 1500 \times 10^{-20} \, \mathrm{erg \, s^{-1} \, cm^{-2}}$, and Johnson-V~$\sim19.0$. In contrast, when comparing data from different nights for NGC\,1433, fainter regions suffer greatly from distinct sky conditions, especially in metallicity measurements, which {range} from $\sim-0.2$ to $\sim+0.2$, strongly affecting the interpretation {of} stellar mass assembly \textcolor{black}{and the origin of the gas (i.e., mergers might bring metal-poor gas, while an in-situ burst of star formation can yield metal-rich stars)}. This study is the first of its kind and highlights the importance of proper sky subtraction, providing insights regarding data limitations and {observational strategies} for future observations with MUSE.}


   \maketitle
   \nolinenumbers

\section{Introduction}

The Multi Unit Spectroscopic Explorer (MUSE)\footnote{https://www.eso.org/sci/facilities/develop/instruments/muse.html}, at the Very Large Telescope (VLT/ESO), is an Integral Field Spectrograph (IFS) and represents a major milestone in the investigation of stellar mass assembly in nearby galaxies -- e.g.,  TIMER (PI: Gadotti; \citealp{gadotti2019time}), PHANGS (PI: Emsellem; \citealp{emsellem2022phangs}), GECKOS (PI: van de Sande; \citealp{2024IAUS..377...27V}), and numerous other programs. It enables the investigation of the spatially and structurally dependent stellar population properties at a spatial sampling of 0.2''~$\times$~0.2'', with a fairly large field of view (60''~$\times$~60'').

One of the crucial steps in achieving relevant results is data reduction, where many interventions must be considered, both to mitigate instrumental imperfections -- such as flats, biases, and darks -- and natural impediments, such as sky emission when observing from the ground. The pipeline provided by ESO\footnote{https://www.eso.org/sci/software/pipelines/muse/} for MUSE is quite complex, since each pointing corresponds to thousands of spectra. To perform sky subtraction, the user can consider the sky around the target (for point-like sources) or a dedicated sky frame, which is the case for extended objects such as nearby galaxies, which can occupy the entire MUSE FoV. In the latter case, it is not clear, however, which dedicated sky pointing strategy results in the best sky subtraction, and different PIs consider different sequences, such as OSO, SOOS, OSOOSO, OOSOO, OOS (where O = object and S = sky). All these strategies assume that the sky background doesn't change significantly between the science frame and corresponding sky frame and, ideally, one would like to take sky frames as few as possible. But this will result in a large time separation between the science and sky frames, and the above assumption may not hold. To date, the impact of this choice on the end products (such as age, metallicity, and $\alpha$-enhancement) is difficult to constrain. \textcolor{black}{Furthermore, many studies that built mosaics with MUSE (i.e., multiple pointings to cover a larger area) found jumps between frames in the values of derived properties (e.g., age, metallicity, dust), which can be at least partially attributed to different sky conditions and subtractions. Many of the dust maps of \cite{emsellem2022phangs} (Figure 23) present jumps between frames, while in \cite{2023A&A...673A.147P}, even with extensive additional processing, many maps still retain mosaic-correlated jumps. To mitigate the impact on dust maps, \cite{2026arXiv260719717K} discarded most of the wavelength regime.}  

In this work, we propose to analyse the impact of sky subtraction when considering different strategies. For this, we consider two case studies: (\textit{i}) sky stability over a period of two hours; and (\textit{ii}) the impact of different sky conditions. We seek to provide the first metrics on how the sky subtraction step might affect unresolved stellar population end products.

\begin{figure*}
        \centering
        \includegraphics[width=0.80\linewidth]{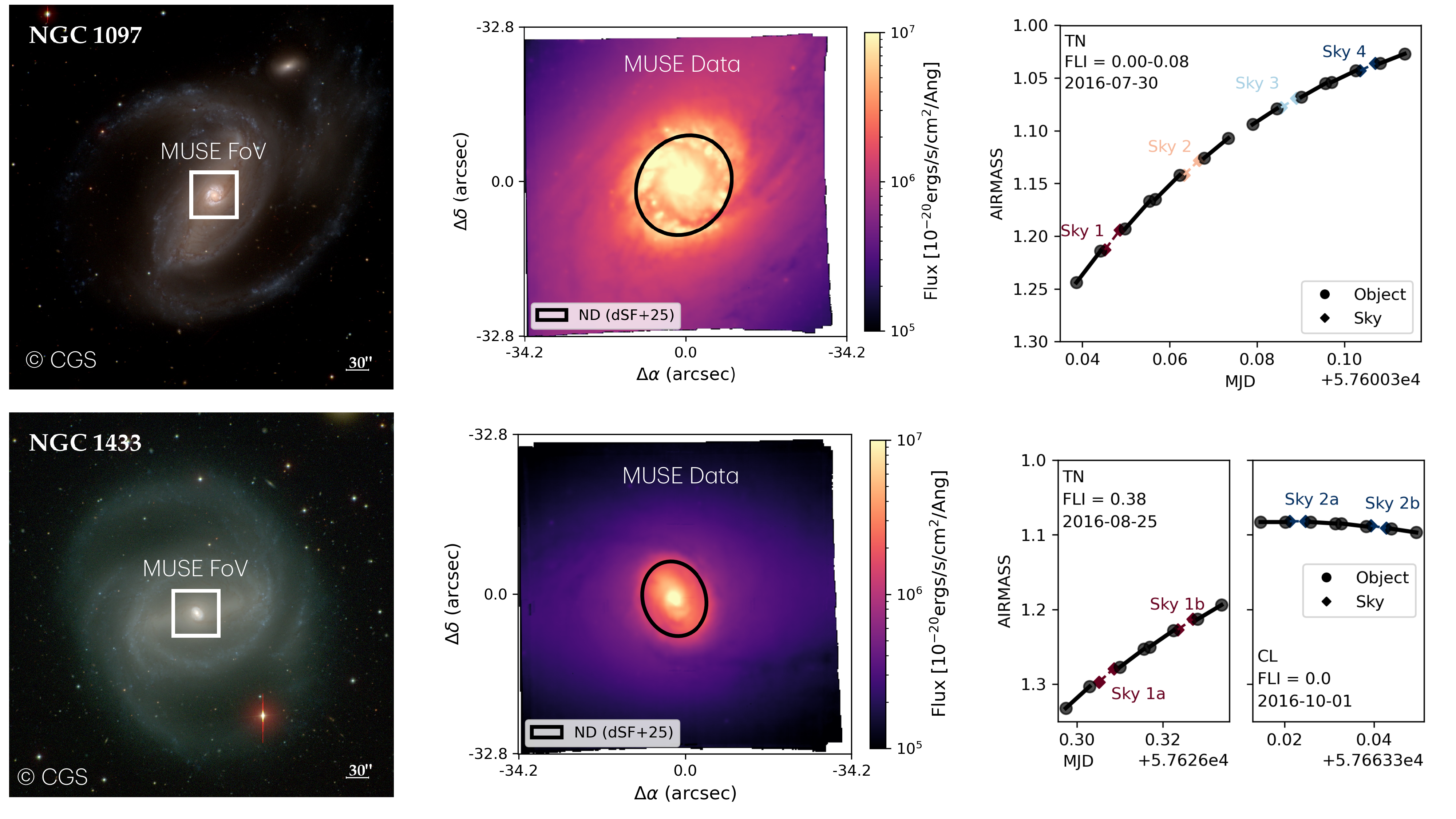}
        \caption{{Sample and observation properties:} top, NGC\,1097, and bottom, NGC\,1433. From left to right, we present the image of the galaxy (from the Carnegie-Irvine Galaxy Survey -- \citealp{ho2011carnegie}), with the MUSE FoV (white square); next, we display the white-light image from MUSE, highlighting the nuclear disc (ND -- solid black contour) size from \cite{de2025bar}; lastly, we display the airmass of the observation as a function of the Modified Julian Date (MJD), differentiating between object and sky pointings (circle and diamonds, respectively). }
        \label{fig_1}
\end{figure*}

\section{Data}
\label{sec_data}

To probe the sky impact on MUSE observations, we opted for the TIMER survey data (PI: Gadotti -- \citealp{gadotti2019time}), which is composed of the central regions of 21 nearby barred galaxies and follows an observational strategy of 2 OBs, totalling 8 pointings on target and 4 on sky (see Figure~\ref{fig_1}), providing enough room for sky-related tests. From the 21 galaxies, we selected two, based on the conditions of their observations: NGC\,1097 and NGC\,1433. The first galaxy had both observing blocks (OBs) observed back-to-back, while the second had one OB observed under less favourable conditions than the other. Finally, both galaxies host a nuclear disc (ND), which is a structure in the central region of the galaxy, built by gas inflows triggered by the stellar bar (e.g., \citealp{athanassoula2005nature, lin2013hydrodynamical, fragkoudi2016close,verwilghen2024simulating}). \cite{de2025bar} measured ND sizes of 1300 and 430~pc hosted by NGC\,1097 and NGC\,1433, respectively. Since the NDs are brighter than their surroundings, they naturally have a higher signal-to-noise ratio (SNR), stretching the SNR range we will manage to probe in this analysis.

The NGC\,1097 observations were split into two consecutive OBs, executed on 30th of July 2016 over approximately two hours. During this time, the airmass decreased from 1.25 to $\sim$1.05. The Moon had a  \textcolor{black}{fraction of light illumination} (FLI) of 0.08 and rose towards the end of the observation. This presents a good opportunity to explore the stability of the sky frames over two hours, with a sky flux variation up to $\sim18\%$ since the start of the observation. In Figure~\ref{fig_spectra_NGC1097_5800}, we display the sky spectra for all sky frames between 4800 and 5800 Angstroms, highlighting how the continuum is not strongly affected, but rather specific lines and regions (see Figure~\ref{fig_spectra_NGC1097_5800_zoom}). On the other hand, NGC\,1433 was observed on two separate nights: the first was 25 August 2016, with FLI~$=0.38$ (technically dark time), thin transparency (TN), and airmass values from $\sim 1.35\text{--}1.2$; and the second was 1 October 2016, with FLI~$=0$, clear transparency, and airmass values of $\sim 1.1$. In Figure~\ref{fig_spectra_NGC1433_5800}, we display the sky spectra of both nights, highlighting how the continuum level is affected ($\sim20\%$), together with specific regions and lines (see Figure~\ref{fig_spectra_NGC1433_5800_zoom}). These distinct sky conditions present an excellent setup to test how atmospheric variations affect derived stellar population properties such as age, $\alpha$-enhancement, and metallicity. 

\section{Methodology}
\label{sec_methodology}

We analyse the impact of sky subtraction on the stellar population products, considering two aspects: (\textit{i}) variation of sky conditions between consecutive OBs, and (\textit{ii}) sky transparency and moon illumination fraction -- in this case, with an angular distance of $\sim70^\circ$.

\subsection{Sky subtraction steps}

For the first case, we consider NGC\,1097 and perform 5 different reductions, changing only the sky subtraction step: the default reduction -- which considers the closest sky frame in the sky subtraction step -- and 4 more reductions, each of which considers a single sky frame to reduce all 8 data cubes at a time (see Figure~\ref{fig_1}). For example, the reduction product we refer to as ``Sky 1'' corresponds to all 8 data cubes being reduced with sky frame 1 in the sky subtraction step. With this, we will be able to assess how far in time and airmass the sky frame can be taken and still retrieve consistent results -- considering a night with FLI~=~0-0.08 and TN sky transparency. And where the differences are originating from.

For the second case, we consider NGC\,1433 and, instead of performing different steps of sky subtraction, we analyse the data cubes of each night separately, with 4 exposures each (following the default reduction), and compare them to the final data cube with 8 exposures. With this, we will compare how different sky conditions might impact the stellar population properties in the same target -- assuming the other steps of data reduction are stable, in this case. It is worth highlighting that, even though the conditions were different, they were within user constraints.

\subsection{Deriving stellar population products}

While we do not mean to provide an exhaustive analysis of the impacts of sky subtraction on stellar population products, we expect to put forward a first case study on the topic. Consequently, we will not present multiple pathways to derive stellar population products, as we are aware that absolute results depend heavily on choices such as the stellar population library, wavelength range, and software of choice, among others (e.g., \citealp{gonccalves2020well}). Instead, we intend to present a relative analysis of how results might vary within the same data set, always considering the identical configuration to derive the stellar population products. 

We follow the procedure presented by the TIMER collaboration (e.g., \citealp{bittner2020inside,de2023disc}): employing the GIST pipeline (\citealp{bittner2019gist}), we selected the wavelength range between 4800 and 5800~$\mathrm{\mathring{A}}$, Voronoi-binned (\citealp{cappellari2003adaptive}) the data to a minimum SNR of 100, and considered the MILES simple stellar population libraries (SSPs -- \citealp{vazdekis2015evolutionary}). In short, the steps performed by GIST and its consecutive products are: first, an unregularised pPXF (\citealp{cappellari2004parametric,cappellari2017improving}) run, deriving stellar kinematic properties such as velocity, velocity dispersion, $h_3$, and $h_4$; second, a pyGANDALF (see details in \citealp{bittner2019gist}) run to fit emission lines with Gaussians  \textcolor{black}{and remove them, producing emission-subtracted} spectra for the next run;  \textcolor{black}{considering the sky-subtracted spectra}, the last step is a regularised pPXF run to derive the weights of each SSP and the average properties such as age, metallicity, and $\alpha$-enhancement. The weights of each SSP are used to reconstruct the spatially resolved star formation history of the galaxy. For more details, we refer the reader to \cite{bittner2020inside} and other TIMER papers. Lastly, we did not measure corresponding uncertainties for stellar population parameters, which can cause differences, especially in lower surface brightness regions, since they present larger uncertainties. 

\section{Results}
\label{sec_results}
In this section, we present the values of age, metallicity, and $\alpha$-enhancement for two case studies presented in this work: NGC\,1097 and NGC\,1433. 

\subsection{NGC\,1097: Sky stability over time}
We take advantage of the NGC\,1097 data set to evaluate how employing different (but still feasible) sky frames can affect the derived ages, metallicities, and $\alpha$-enhancements. The median surface brightness of the sky white-light images varies from $\sim2200-2600$ $\times 10^{-20} \, \mathrm{erg \, s^{-1} \, cm^{-2}}\, \mathrm{ang}^{-2}$. Compared to the properties of a census of the 2024 sky frames under different sky characteristics (private communication, ESO), this variation over a period of two hours is standard ($\approx 300$ $\times 10^{-20} \, \mathrm{erg \, s^{-1} \, cm^{-2}}\, \mathrm{ang}^{-2}$), while the values are on the higher end -- possibly due to proximity to the morning twilight.

\begin{figure}
        \centering
        \includegraphics[width=\linewidth]{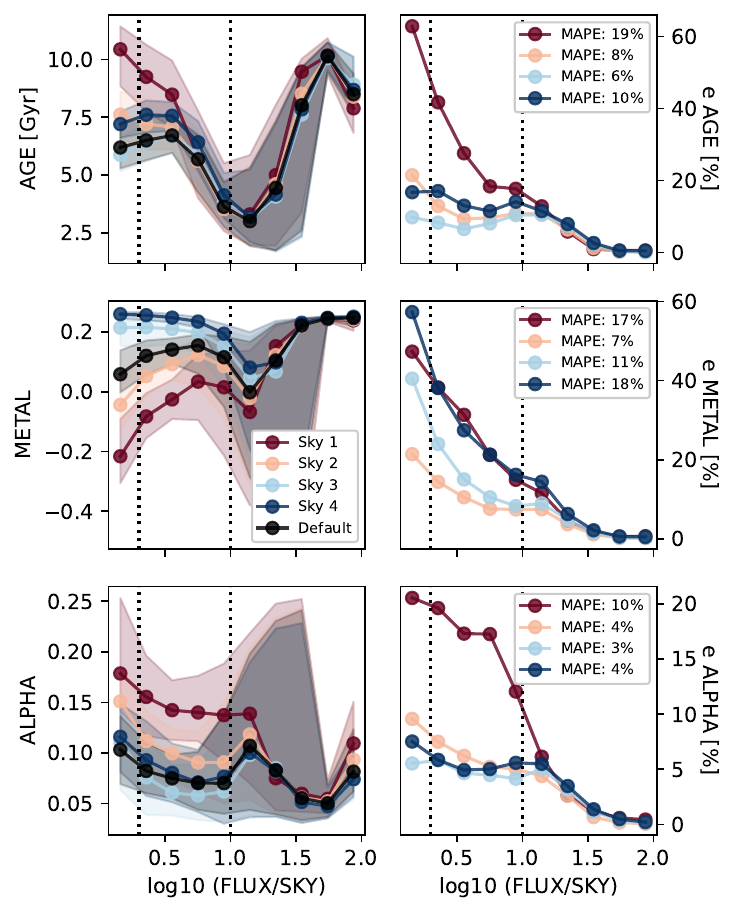}
        \caption{{Stellar population properties with respect to relative flux of NGC\,1097:} In the left column, we display the profiles of age (top), metallicity (middle), and $\alpha$-enhancement (bottom) with respect to relative data flux (see Section~\ref{sec_methodology}), with corresponding $1\sigma$ distributions. In the right column, we display the respective error (see Equation~\ref{eq_1}). The vertical dotted lines serve as eye-guidance for 2  and 10 $\times$ brighter fluxes with respect to the sky background.}
        \label{fig_FLUXRatio_NGC1097}
\end{figure}

To analyse the dependency of the results on the chosen sky subtraction, we analysed the 5 different reductions (see Section~\ref{sec_methodology} for details) in Figure~\ref{fig_FLUXRatio_NGC1097}. We display median binned profiles of the stellar properties with respect to the relative flux of the target (defined as the ratio between the white-light images of the reduced data and the sky frame used in the reduction), in 10 bins. We also present the relative error in comparison with the default reduction in each bin, defined as:

\begin{equation}
    \mathrm{error = 100\times \left\lvert \frac{SKY_*-DEFAULT}{DEFAULT} \right\lvert}
    \label{eq_1}
\end{equation}

Additionally, we measured the Mean Absolute Percentage Error (MAPE) for each profile as follows:

\begin{equation}
    \mathrm{MAPE} = \frac{100}{n}\times \sum_n \mathrm{ \left\lvert\frac{SKY_*-DEFAULT}{DEFAULT} \right\lvert}
    \label{eq_2}
\end{equation}

\noindent where $n$ is the number of bins, which is 10. It is worth highlighting that the error (Equation~\ref{eq_1}) is measured spatially, and the metallicity and the $\alpha$-enhancement are linearised for it. 

\begin{figure}[!h]
        \centering
        \includegraphics[width=1\linewidth]{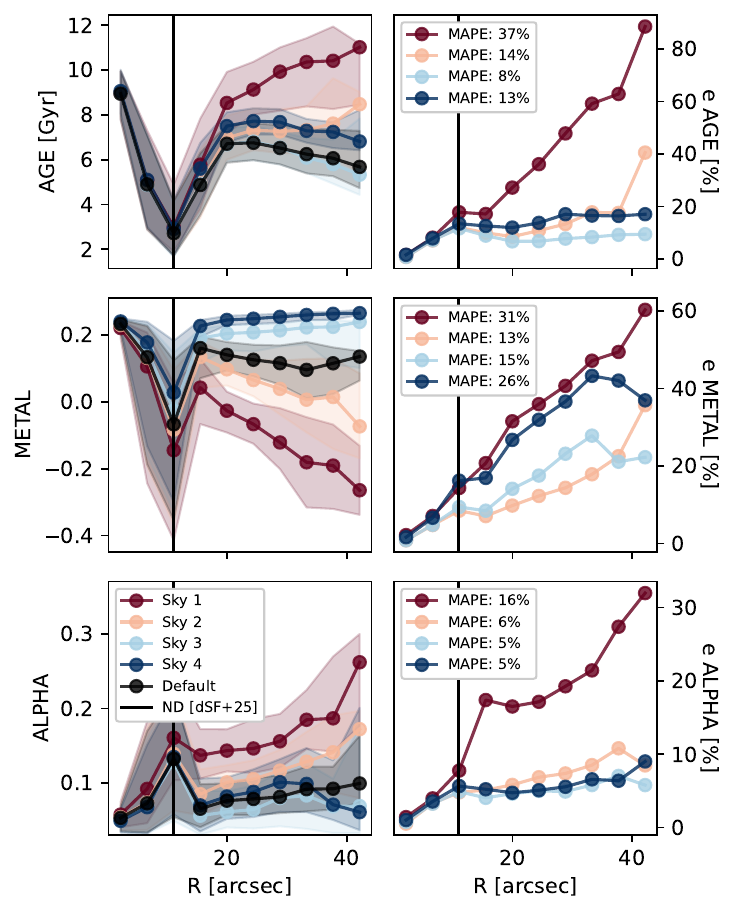}
        \caption{{Radial profiles of stellar population properties:} similar to Figure~\ref{fig_R_NGC1433}, we display the median radial profiles of age (top), metallicity (middle), and $\alpha$-enhancement (bottom) for the different sky subtractions in the left column (see Section~\ref{sec_methodology}), with corresponding $1\sigma$ distributions. Additionally, we display the relative error of the different sky reductions with respect to the default sky subtraction in the right column. The vertical solid line corresponds to the nuclear disc (ND) size from \cite{de2025bar}.}
        \label{fig_R_NGC1097}
\end{figure}

\begin{figure}
        \centering
        \includegraphics[width=1\linewidth]{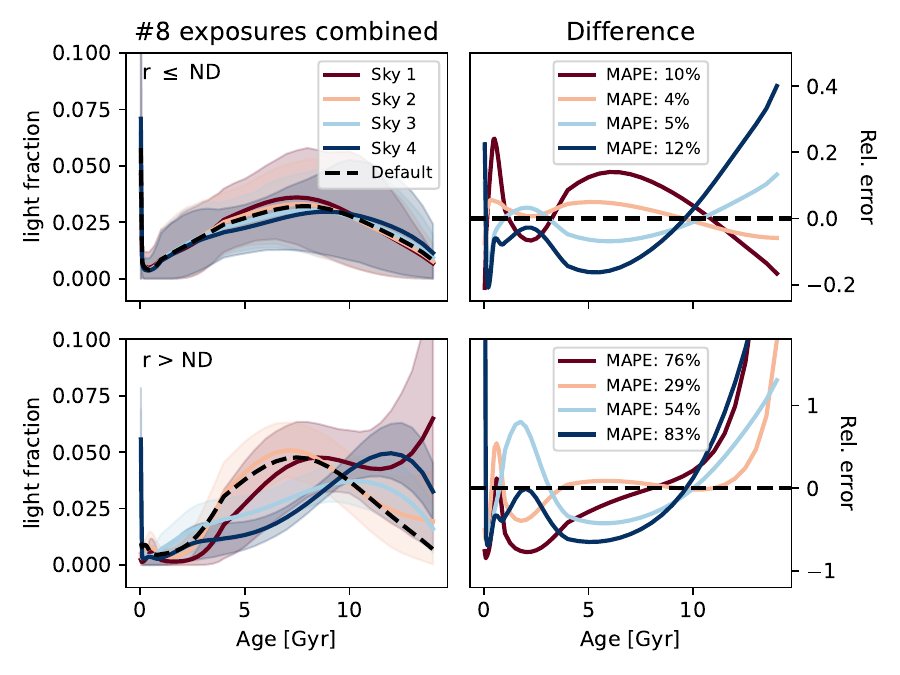}
        \caption{{SFH from different sky subtractions:} We display the stellar population weights (light fraction) with respect to stellar population age inside the nuclear disc (top) and outside the nuclear disc (bottom) in the left column -- i.e., we measured the median SFH for the bins inside and outside the nuclear disc, with the $1\sigma$ distribution in shaded regions. In the right column, we display the relative error (similar to Equation~\ref{eq_1}, but without considering the absolute values of the differences or percentages) with respect to the default reduction, along with the corresponding MAPE (Equation~\ref{eq_2}). While the impact of sky subtraction seems to follow a trend with airmass for the star formation history inside the nuclear disc (top-right), this does not apply to the regions outside the nuclear disc, which are fainter and more susceptible to other variables.}
        \label{fig_SFH_NGC1097}
\end{figure}

\begin{figure}
        \centering
        \includegraphics[width=\linewidth]{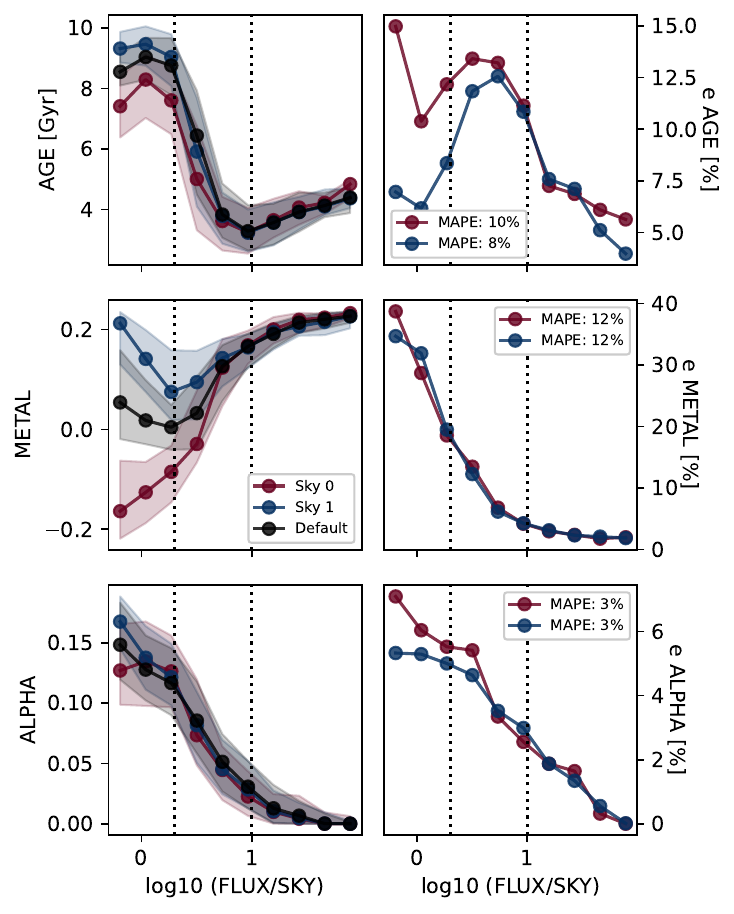}
        \caption{{Stellar population properties with respect to relative flux of NGC\,1433:} Similar to Figure~\ref{fig_FLUXRatio_NGC1097}, but considering the NGC\,1433 data from different nights (see Figure~\ref{fig_1}).}
        \label{fig_FLUXRatio_NGC1433}
\end{figure}

\begin{figure}
        \centering
        \includegraphics[width=1\linewidth]{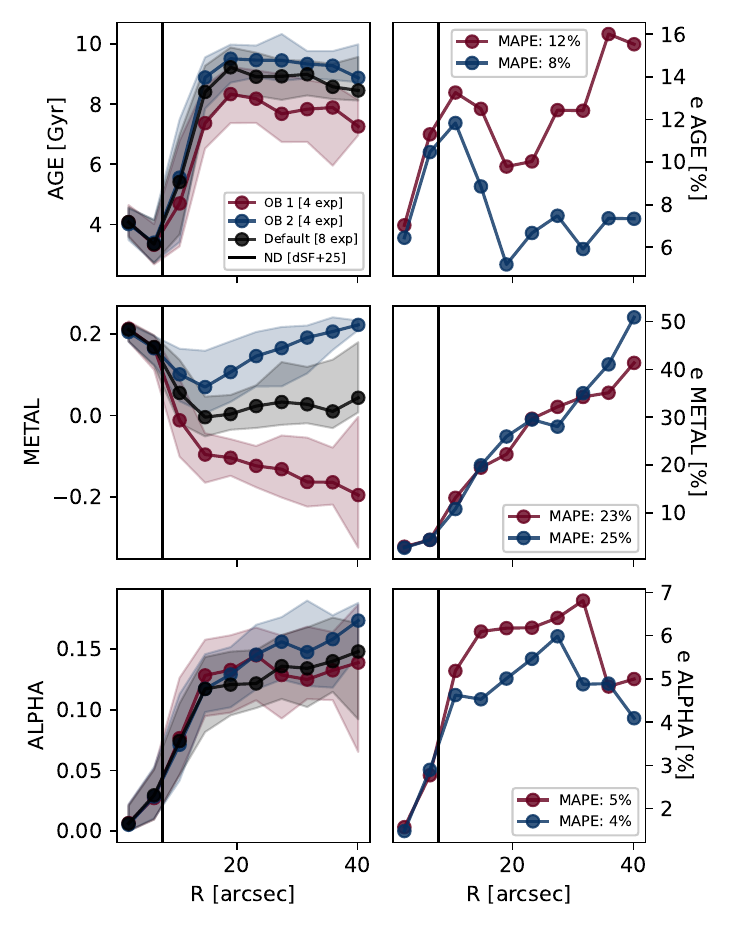}
        \caption{{Radial profiles of stellar population properties:} we display median and $1\sigma$ values of age, metallicity, and $\alpha$-enhancement with respect to radius (left column) and the corresponding relative errors (Equation~\ref{eq_1} -- right column). The default reduction combining 8 exposures is shown with black lines and markers, while the OBs for single nights (combining 4 exposures) are in red and blue for the first and second nights, respectively. Lastly, the nuclear disc radius is indicated by the solid black line (\citealp{de2025bar}). As expected, inside the nuclear disc, the results seem to be quite consistent. Outside the nuclear disc, however, the differences can be substantial, especially in metallicity -- nevertheless, the MAPE values for metallicity are stretched because the default values are close to zero.}
        \label{fig_R_NGC1433}
\end{figure}

Of the three properties, age and $\alpha$-enhancement are the ones that seem to be the least affected by different sky reductions, with low MAPE values ranging from $\sim 4\text{--}19\%$ -- not surprisingly, the intermediate skies 2 and 3 have the smallest MAPE values (see Appendix~\ref{appendix_b} for sky spectra). For both properties, skies 1-3 perform better, with errors below $20\%$ and $5\%$ for age and metallicity, respectively. The metallicity, on the other hand, consistently reaches higher error values, especially for Skies 1, while maintaining the MAPE values below $20\%$. For regions with relative fluxes above 10 (corresponding to values above $\sim 1200 \times 10^{-20} \, \mathrm{erg \, s^{-1} \, cm^{-2}}$), the sky reduction of choice seems almost irrelevant, yielding very consistent results, both for the median and $1\sigma$ values. Finally, for relative fluxes between $2-10$, the skies 2 and 3 (intermediate skies) perform with error below $20\%$, while 1 and 4 reach higher values (assuming the default reduction represents the best option available). Since MILES is not equipped with stellar populations younger than 30~Myr, it can yield solutions of young, metal-poor stellar populations coinciding with star-forming regions. An extensive analysis of the templates of choice is beyond the scope of this paper, but in Appendix~\ref{appendix_a} we repeat the analysis masking these regions, finding similar results. Furthermore, \cite{gonccalves2026model} presented the dependency of the results with respect to the stellar library of choice and, even though the absolute numbers vary, most libraries yield similar general trends. 

Lastly, to understand the origin of these differences, we analysed the resulting spectra in Figure~\ref{fig_spectra_NGC1097_5800}. Among our findings, we noticed that the sky line [NI] $\lambda5200$ falls precisely in the redshifted Mg$_b$ bandpass and, if not properly accounted for  \textcolor{black}{(especially in the measurement of Lick indices)}, can influence the final stellar population properties (see Appendix~\ref{appendix_b} for more details). This poses a particular challenge for studies of galaxies with redshifts between 0.0014 and 0.0077.

\subsection{Structure-related analysis }

To further investigate the spatial dependence of the results presented in Figure~\ref{fig_FLUXRatio_NGC1097}, we analyse the results of the different sky subtractions for NGC\,1097 radially in Figure~\ref{fig_R_NGC1097}. Consistent with Figure~\ref{fig_FLUXRatio_NGC1097}, we find that age, metallicity, and $\alpha$-enhancement are not strongly affected by the choice of sky subtraction inside the nuclear disc, where we expect the highest SNRs. On the other hand, outside the nuclear disc, the differences can be much more significant, with higher MAPE values. Skies 1 and 4 (the first and last exposures) result in the largest differences in stellar properties, while skies 2 and 3 yield results closer to the default reduction. In this case, the higher impact on end-products is outside the nuclear disc and, analysing the median spectra of a ring just outside it -- width of 2'', we find the median white-image flux of $\sim 1500 \times 10^{-20} \, \mathrm{erg \, s^{-1} \, cm^{-2}}$ -- and estimate a magnitude of $\sim19$ in the V-band of the Johnson system (reference point Vega).

Furthermore, we analysed the median SFHs inside and outside the nuclear disc in Figure~\ref{fig_SFH_NGC1097}, investigating which particular ages might be affected. Inside the nuclear disc, the SFH differences are minimal, reaching a maximum MAPE value of $12\%$ with the final sky frame (i.e., the one with the lowest airmass value). Skies 2 and 3 perform better once again, with MAPE values of $4\text{--}5\%$. The differences among specific stellar populations also seem to follow a pattern: skies with higher airmass values tend to yield younger stellar populations than those with lower airmass values. That is, if the sky frame has a higher airmass than the target (and is unrepresentative), it will over-subtract the atmospheric signature. The opposite is true when the sky frame has a lower airmass than the target, under-subtracting the atmosphere. To confirm this trend, it would be necessary to perform a systematic analysis of multiple galaxies under similar conditions, which is beyond the scope of this letter. The star formation history outside the nuclear disc, on the other hand, does not follow the same patterns: there is no clear trend with airmass or stellar population, illustrating how complex this process becomes for fainter regions.

\subsection{NGC\,1433: Observational constraints consistency}

Oftentimes, it is the case that OBs for the same target are observed on different nights, under different sky conditions (within the PI's constraints). In this section, we consider the different observations of NGC\,1433 and compare the default reduction with the individual data cubes from each night. 

Performing a similar analysis for NGC\,1433 in Figure~\ref{fig_FLUXRatio_NGC1433}, we find, for relative fluxes above 10, encouraging results, with relative errors $\sim0-10\%$. Moreover, for relative fluxes between 2 and 10, all properties perform fairly well, with relative errors below $20\%$. Lastly, for relative fluxes below 2, metallicity seems to be the most affected property, varying from $-0.2$ to $+0.2$, which inherently affects the interpretation around stellar mass assembly. It is important to highlight, however, that the spatial resolution when one bins the data to a specific signal-to-noise ratio (e.g., Voronoi bins, \citealp{cappellari2003adaptive}) can be affected by the choice of combining 4 or 8 exposures, especially in the regions outside the nuclear disc. 

In Figure~\ref{fig_R_NGC1433}, we display the median radial profiles of the different stellar population properties (age, metallicity, and $\alpha$-enhancement) for the different OBs, with the corresponding $1\sigma$ distributions (shaded regions). As in NGC\,1097, the results are very consistent between nights inside the nuclear disc region for all properties. Outside, however, this is not the case, following the expectation related to fainter regions. Compared with the first night, the OB from the second night (i.e., better sky conditions) shows generally similar ages and $\alpha$-enhancement (within $1\sigma$) but higher metallicity values. This can contribute to the “jumps” in stellar property values witnessed in mosaics, which are usually present in fainter regions (noticeably fainter than the regions presented here). When comparing the nights directly, the differences are even greater -- in the case of metallicity, the values are negative for one night and positive for the other, strongly affecting the mass assembly interpretation -- often, metal-poor stars are connected with infalling gas from interactions/mergers, while metal-rich stars are connected to in-situ star formation (where very high metallicities can be associated with an intense and fast enrichment star-formation episode).

\section{Conclusions and Summary}

When observing extended objects with MUSE, one needs to include sky pointings in their observing plan for proper data reduction. The impact of the chosen sky frame on the final product, however, is not well constrained, depending on brightness and the science case, which potentially impacts the efficiency of telescope time usage. Furthermore, the same target often needs to be observed on different nights, under different sky conditions, within a threshold of acceptance defined by the user. In this letter, we aim to provide an insightful analysis of how sky subtraction can affect stellar population properties, considering two case studies. We summarise our main findings below:

\begin{itemize}

    \item We find that when considering sky frames taken within a period of two hours for NGC\,1097 (with TN condition and airmass values between 1.25 and 1.05), fainter regions are the most affected by different sky subtractions, while regions with relative fluxes above 10 are not strongly affected. In this case, the threshold corresponds to flux values above $\sim 1200 \times 10^{-20} \, \mathrm{erg \, s^{-1} \, cm^{-2}}$, and $\sim18$ mag (Johnson-V, Vega). We cannot, within this letter, generalise these values, but they can be considered a reference, depending on the scientific case and relative brightness with the sky;
    
    \item  \textcolor{black}{When considering stellar structures of NGC\,1097, the results within the nuclear disc do not vary strongly, depending on the sky-subtraction choice. While outside the nuclear disc, the impact is much more noticeable. When it comes to the star formation histories within and outside the nuclear disc, the trend with sky-subtraction within the nuclear disc is clear, where under representative skies (lower airmass values) yield younger stellar populations than over representative skies (higher airmass values). To confirm this trend, it is necessary to repeat the analysis presented here in a systematic way for a larger sample;}

    \item When analysing OBs of NGC\,1433 from different nights, we see consistent values for relative fluxes above 10, corresponding to the nuclear disc region. Furthermore, in regions with lower relative fluxes, metallicity seems to be the most affected, changing from $\sim -0.2$ to $\sim +0.2$ in those regions, strongly affecting the interpretation of mass assembly  \textcolor{black}{(concerning the characteristics of the gas that originated these stars)} for this specific case. Radially, this coincides with the nuclear disc as well;

    \item  \textcolor{black}{We repeated the analysis for NGC\,1097 but masked young metal-poor regions that can be a result of limitations related to the stellar library of choice, finding no difference in the results};
    
    \item We investigate which features in the spectra might cause these changes, finding that the Mg$_b$ bandpass is strongly affected by the sky line [NI] $\lambda$5200, which impacts the stellar population results for galaxies within $0.0014\leq\mathrm{z}\leq0.0077$ if the sky is not properly removed. 
    
\end{itemize}

While this is the first study of its kind, confirming the values and conclusions found here requires reproducing similar analyses on a larger data set. We do expect, however, these results to provide MUSE users with insights regarding observing strategies, data analysis, and unresolved stellar population modelling, even if still within specific constraints (e.g., the MILES library, specific wavelength ranges, pPXF, etc.).

\begin{acknowledgements}
We thank the great report from the referee Eva Sextl, which we believe improved the results in this paper. CdSF thanks Amelia Fraser-McKelvie for the many insightful conversations on the subject, including the spectra investigation. 
\end{acknowledgements}

\bibliographystyle{bibtex/aa} 
\bibliography{bibliography} 

@article{cappellari2003adaptive,
  title={Adaptive spatial binning of integral-field spectroscopic data using Voronoi tessellations},
  author={Cappellari, Michele and Copin, Yannick},
  journal={MNRAS},
  volume={342},
  number={2},
  pages={345--354},
  year={2003},
  publisher={Blackwell Science Ltd Oxford, UK}
}

@article{cappellari2004parametric,
  title={Parametric recovery of line-of-sight velocity distributions from absorption-line spectra of galaxies via penalized likelihood},
  author={Cappellari, Michele and Emsellem, Eric},
  journal={PASP},
  volume={116},
  number={816},
  pages={138},
  year={2004},
  publisher={IOP Publishing}
}

@article{athanassoula2005nature,
  title={On the nature of bulges in general and of box/peanut bulges in particular: input from N-body simulations},
  author={Athanassoula, Evangelista},
  journal={MNRAS},
  volume={358},
  number={4},
  pages={1477--1488},
  year={2005},
  publisher={The Royal Astronomical Society}
}

@article{ho2011carnegie,
  title={The carnegie-irvine galaxy survey. I. Overview and atlas of optical images},
  author={Ho, Luis C and Li, Zhao-Yu and Barth, Aaron J and Seigar, Marc S and Peng, Chien Y},
  journal={ApJS},
  volume={197},
  number={2},
  pages={21},
  year={2011},
  publisher={IOP Publishing}
}

@article{lin2013hydrodynamical,
  title={Hydrodynamical simulations of the barred spiral galaxy NGC 1097},
  author={Lin, Lien-Hsuan and Wang, Hsiang-Hsu and Hsieh, Pei-Ying and Taam, Ronald E and Yang, Chao-Chin and Yen, David CC},
  journal={ApJ},
  volume={771},
  number={1},
  pages={8},
  year={2013},
  publisher={IOP Publishing}
}

@article{vazdekis2015evolutionary,
  title={Evolutionary stellar population synthesis with MILES--II. Scaled-solar and $\alpha$-enhanced models},
  author={Vazdekis, A and Coelho, P and Cassisi, Santi and Ricciardelli, E and Falc{\'o}n-Barroso, J and S{\'a}nchez-Bl{\'a}zquez, Patricia and Barbera, F La and Beasley, MA and Pietrinferni, Adriano},
  journal={MNRAS},
  volume={449},
  number={2},
  pages={1177--1214},
  year={2015},
  publisher={Oxford University Press}
}

@article{fragkoudi2016close,
  title={A close look at secular evolution: boxy/peanut bulges reduce gas inflow to the central kiloparsec},
  author={Fragkoudi, F and Athanassoula, E and Bosma, A},
  journal={MNRAS},
  volume={462},
  number={1},
  pages={L41--L45},
  year={2016},
  publisher={The Royal Astronomical Society}
}

@article{cappellari2017improving,
  title={Improving the full spectrum fitting method: Accurate convolution with Gauss--Hermite functions},
  author={Cappellari, Michele},
  journal={MNRAS},
  volume={466},
  number={1},
  pages={798--811},
  year={2017},
  publisher={Oxford University Press}
}

@article{bittner2019gist,
  title={The GIST pipeline: A multi-purpose tool for the analysis and visualisation of (integral-field) spectroscopic data},
  author={Bittner, Adrian and Falc{\'o}n-Barroso, J and Nedelchev, B and Dorta, A and Gadotti, DA and Sarzi, M and Molaeinezhad, A and Iodice, ENRICHETTA and Rosado-Belza, D and de Lorenzo-C{\'a}ceres, A and others},
  journal={A\&A},
  volume={628},
  pages={A117},
  year={2019},
  publisher={EDP Sciences}
}

@article{gadotti2019time,
  title={Time Inference with MUSE in Extragalactic Rings (TIMER): properties of the survey and high-level data products},
  author={Gadotti, Dimitri A and S{\'a}nchez-Bl{\'a}zquez, Patricia and Falc{\'o}n-Barroso, Jes{\'u}s and Husemann, Bernd and Seidel, Marja K and P{\'e}rez, Isabel and de Lorenzo-C{\'a}ceres, Adriana and Martinez-Valpuesta, Inma and Fragkoudi, Francesca and Leung, Gigi and others},
  journal={MNRAS},
  volume={482},
  number={1},
  pages={506--529},
  year={2019},
  publisher={Oxford University Press}
}

@article{gonccalves2020well,
  title={How well can we determine ages and chemical abundances from spectral fitting of integrated light spectra?},
  author={Gon{\c{c}}alves, Geraldo and Coelho, Paula and Schiavon, Ricardo and Usher, Christopher},
  journal={MNRAS},
  volume={499},
  number={2},
  pages={2327--2339},
  year={2020},
  publisher={Oxford University Press}
}

@article{bittner2020inside,
  title={Inside-out formation of nuclear discs and the absence of old central spheroids in barred galaxies of the TIMER survey},
  author={Bittner, Adrian and S{\'a}nchez-Bl{\'a}zquez, Patricia and Gadotti, Dimitri A and Neumann, Justus and Fragkoudi, Francesca and Coelho, Paula and de Lorenzo-C{\'a}ceres, Adriana and Falc{\'o}n-Barroso, Jes{\'u}s and Kim, Taehyun and Leaman, Ryan and others},
  journal={A\&A},
  volume={643},
  pages={A65},
  year={2020},
  publisher={EDP Sciences}
}

@article{emsellem2022phangs,
  title={The PHANGS-MUSE survey-Probing the chemo-dynamical evolution of disc galaxies},
  author={Emsellem, Eric and Schinnerer, Eva and Santoro, Francesco and Belfiore, Francesco and Pessa, Ismael and McElroy, Rebecca and Blanc, Guillermo A and Congiu, Enrico and Groves, Brent and Ho, I-Ting and others},
  journal={A\&A},
  volume={659},
  pages={A191},
  year={2022},
  publisher={EDP Sciences}
}

@article{de2023disc,
  title={Disc galaxies are still settling-Discovery of the smallest nuclear discs and their young stellar bars},
  author={de S{\'a}-Freitas, Camila and Gadotti, Dimitri A and Fragkoudi, Francesca and Coccato, Lodovico and Coelho, Paula and de Lorenzo-C{\'a}ceres, Adriana and Falc{\'o}n-Barroso, Jes{\'u}s and Kolcu, Tutku and Mart{\'\i}n-Navarro, Ignacio and Mendez-Abreu, Jairo and others},
  journal={A\&A},
  volume={678},
  pages={A202},
  year={2023},
  publisher={EDP Sciences}
}

@article{verwilghen2024simulating,
  title={Simulating nearby disc galaxies on the main star formation sequence I. Bar formation and the building of the central gas reservoir},
  author={Verwilghen, Pierrick and Emsellem, Eric and Renaud, Florent and Valentini, Milena and Sun, Jiayi and Jeffreson, Sarah and Klessen, Ralf S and Sormani, Mattia C and Barnes, Ashley and Dolag, Klaus and others},
  journal={A\&A},
  volume={687},
  pages={53},
  year={2024}
}

@article{de2025bar,
  title={Bar ages derived for the first time in nearby galaxies-Insights into secular evolution from the TIMER sample},
  author={de S{\'a}-Freitas, Camila and Gadotti, Dimitri A and Fragkoudi, Francesca and Coelho, Paula and de Lorenzo-C{\'a}ceres, Adriana and Falc{\'o}n-Barroso, Jes{\'u}s and S{\'a}nchez-Bl{\'a}zquez, Patricia and Kim, Taehyun and Mendez-Abreu, Jairo and Neumann, Justus and others},
  journal={A\&A},
  volume={698},
  pages={A5},
  year={2025},
  publisher={EDP Sciences}
}

@article{gonccalves2026model,
  title={Model-dependent metallicities of young stellar populations in TIMER galaxies},
  author={Gon{\c{c}}alves, Geraldo and Coelho, Paula RT and Silva-Lima, Luiz A and Martins, Lucimara and Martig, Marie and Gadotti, Dimitri A and de S{\'a}-Freitas, Camila and Mendez-Abreu, Jairo and S{\'a}nchez-Bl{\'a}zquez, Patricia and Falc{\'o}n-Barroso, Jes{\'u}s and others},
  journal={Monthly Notices of the Royal Astronomical Society},
  pages={stag1939},
  year={2026},
  publisher={Oxford University Press}
}

@article{verro2022x,
  title={The X-shooter Spectral Library (XSL): data release 3},
  author={Verro, Kristiina and Trager, SC and Peletier, RF and Lan{\c{c}}on, A and Gonneau, A and Vazdekis, A and Prugniel, Philippe and Chen, Y-P and Coelho, PRT and S{\'a}nchez-Bl{\'a}zquez, P and others},
  journal={Astronomy \& Astrophysics},
  volume={660},
  pages={A34},
  year={2022},
  publisher={EDP Sciences}
}

@article{verro2022modelling,
  title={Modelling simple stellar populations in the near-ultraviolet to near-infrared with the X-shooter Spectral Library (XSL)},
  author={Verro, Kristiina and Trager, SC and Peletier, RF and Lan{\c{c}}on, A and Arentsen, A and Chen, Y-P and Coelho, PRT and Dries, M and Falc{\'o}n-Barroso, J and Gonneau, A and others},
  journal={Astronomy \& Astrophysics},
  volume={661},
  pages={A50},
  year={2022},
  publisher={EDP Sciences}
}

@article{leitherer1999starburst99,
  title={Starburst99: synthesis models for galaxies with active star formation},
  author={Leitherer, Claus and Schaerer, Daniel and Goldader, Jeffrey D and Delgado, Rosa M Gonz{\'a}lez and Robert, Carmelle and Kune, Denis Foo and De Mello, Du{\'\i}lia F and Devost, Daniel and Heckman, Timothy M},
  journal={The Astrophysical Journal Supplement Series},
  volume={123},
  number={1},
  pages={3--40},
  year={1999}
}

@article{millan2021hr,
  title={HR-pypopstar: high-wavelength-resolution stellar populations evolutionary synthesis model},
  author={Mill{\'a}n-Irigoyen, I and Moll{\'a}, M and Cervi{\~n}o, M and Ascasibar, Y and Garc{\'\i}a-Vargas, ML and Coelho, PRT},
  journal={Monthly Notices of the Royal Astronomical Society},
  volume={506},
  number={4},
  pages={4781--4799},
  year={2021},
  publisher={Oxford University Press}
}

@INPROCEEDINGS{2024IAUS..377...27V,
       author = {{van de Sande}, J. and {Fraser-McKelvie}, A. and {Fisher}, D.~B. and {Martig}, M. and {Hayden}, M.~R. and {Geckos Survey Collaboration}},
        title = "{GECKOS: Turning galaxy evolution on its side with deep observations of edge-on galaxies}",
    booktitle = {Early Disk-Galaxy Formation from JWST to the Milky Way},
         year = 2024,
       editor = {{Tabatabaei}, Fatemeh and {Barbuy}, Beatriz and {Ting}, Yuan-Sen},
       series = {IAU Symposium},
       volume = {377},
        month = jan,
        pages = {27-33},
          doi = {10.1017/S1743921323001138},
archivePrefix = {arXiv},
       eprint = {2306.00059},
 primaryClass = {astro-ph.GA},
       adsurl = {https://ui.adsabs.harvard.edu/abs/2024IAUS..377...27V}
}

@ARTICLE{2023A&A...673A.147P,
       author = {{Pessa}, I. and {Schinnerer}, E. and {Sanchez-Blazquez}, P. and {Belfiore}, F. and {Groves}, B. and {Emsellem}, E. and {Neumann}, J. and {Leroy}, A.~K. and {Bigiel}, F. and {Chevance}, M. and {Dale}, D.~A. and {Glover}, S.~C.~O. and {Grasha}, K. and {Klessen}, R.~S. and {Kreckel}, K. and {Kruijssen}, J.~M.~D. and {Pinna}, F. and {Querejeta}, M. and {Rosolowsky}, E. and {Williams}, T.~G.},
        title = "{Resolved stellar population properties of PHANGS-MUSE galaxies}",
      journal = {\aap},
         year = 2023,
        month = may,
       volume = {673},
          eid = {A147},
        pages = {A147},
          doi = {10.1051/0004-6361/202245673},
archivePrefix = {arXiv},
       eprint = {2303.13676},
 primaryClass = {astro-ph.GA},
       adsurl = {https://ui.adsabs.harvard.edu/abs/2023A&A...673A.147P}
}

@ARTICLE{2026arXiv260719717K,
       author = {{Kang}, Xiaoyu and {Sextl}, Eva and {Kudritzki}, Rolf-Peter and {Yesuf}, Hassen M. and {Zhang}, Fenghui and {Chang}, Ruixiang and {Li}, Xiejin and {Han}, Yunkun},
        title = "{When Galaxies Refuel: Evolution of the Perturbed Spiral Galaxy NGC1385}",
      journal = {arXiv e-prints},
         year = 2026,
        month = jul,
          eid = {arXiv:2607.19717},
        pages = {arXiv:2607.19717},
          doi = {10.48550/arXiv.2607.19717},
archivePrefix = {arXiv},
       eprint = {2607.19717},
 primaryClass = {astro-ph.GA},
       adsurl = {https://ui.adsabs.harvard.edu/abs/2026arXiv260719717K}
}

\begin{appendix}
 \textcolor{black}{\section{The impact of star-forming regions in NGC\,1097}}
\label{appendix_a}

\begin{figure}[!h]
        \centering
        \includegraphics[width=1\linewidth]{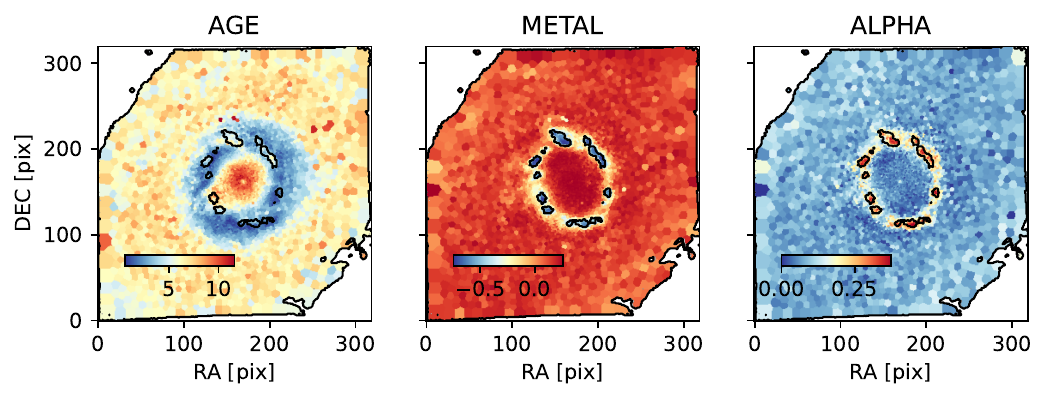}
        \caption{{Stellar population property maps of NGC\,1097:} We present the spatial distribution of age (left), metallicity (middle), and $\alpha-$enhancement (right) of NGC\,1097, for the default reduction run. We also present the regions that were masked in Appendix~\ref{appendix_a}, where metallicity is below -0.4 -- black contours.}
        \label{fig_NGC1097_MPmask}
\end{figure}

Different stellar libraries cover a specific range of properties, such as age, metallicity, and $\alpha-$enhancement values. In this work, we employed MILES (\citealp{vazdekis2015evolutionary}), with ages between 0.03 and 14 Gyr, metallicity between -0.96 and 0.4, and $\alpha-$enhancement of 0 and 0.4. It is known, however, that in the face of young star-forming regions, the MILES library might raise unphysical solutions, since it does not include regions younger than 30 Myr (e.g., \citealp{bittner2020inside,gonccalves2020well}). As a result, it produces extremely metal-poor and young regions that are not expected. This is the case for some regions within the star-forming ring in NGC\,1097 (see example in Fig.~\ref{fig_NGC1097_MPmask}, black contours), and we will address how this affects our analysis in this Appendix.

\begin{figure*}
        \centering
        \includegraphics[width=0.45\linewidth]{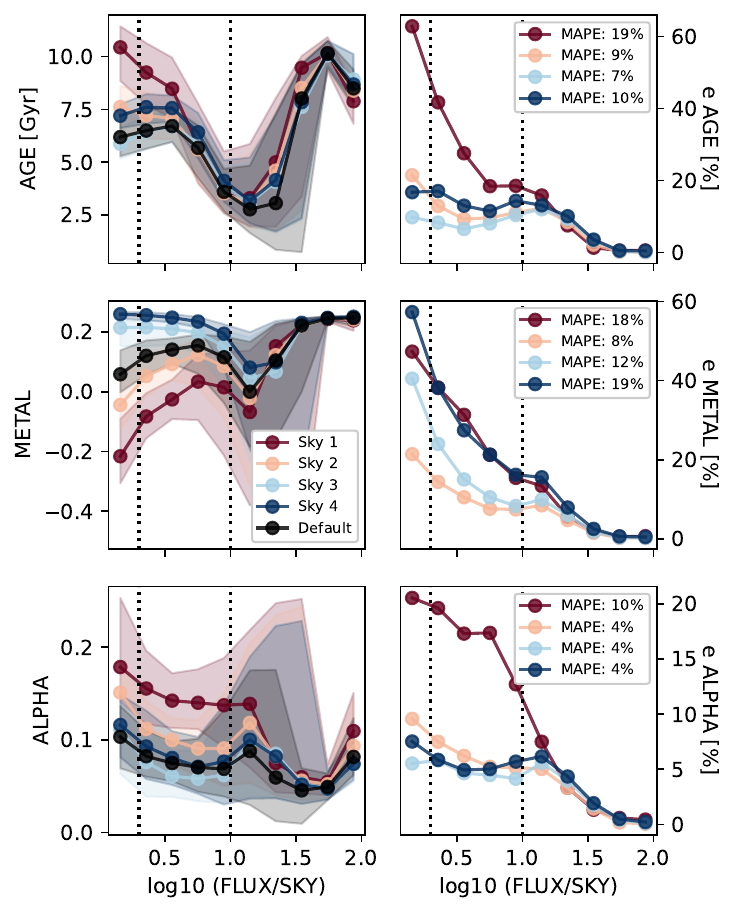}
        \includegraphics[width=0.45\linewidth]{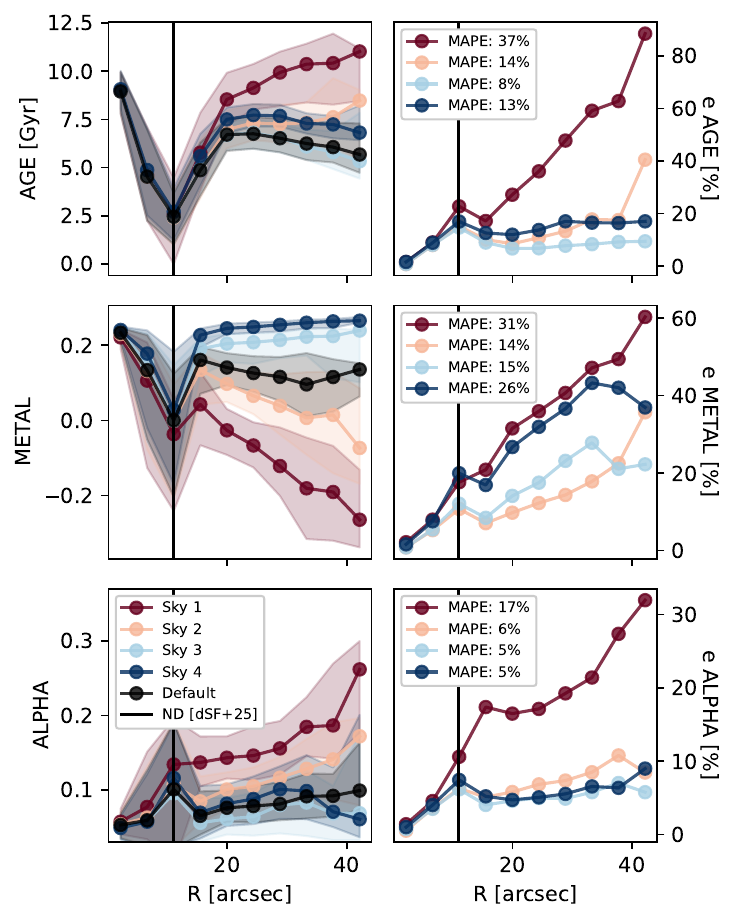}
        \caption{Similar to Figures~\ref{fig_FLUXRatio_NGC1097} (left) and \ref{fig_R_NGC1097} (right), but masking metal-poor regions highlighted in Figure~\ref{fig_NGC1097_MPmask}. The results are not strongly affected.}
        \label{fig_FLUXRatio_NGC1097_MPmask}
\end{figure*}

Recently, \cite{gonccalves2026model} performed a detailed analysis of how the choice of stellar library might affect the final stellar population properties solutions. They considered 11 centrally star-forming galaxies from the TIMER sample (\citealp{gadotti2019time,bittner2020inside}), specifically because \cite{bittner2020inside} found regions of young, metal-poor stars when considering the MILES set of templates. In summary, \cite{gonccalves2026model} derived stellar population properties considering four stellar libraries, varying between empirical and model-based ones: MILES (\citealp{vazdekis2015evolutionary}), X-Shooter (XSL - \citealp{verro2022x,verro2022modelling}), Starburst99 (S99 - \citealp{leitherer1999starburst99}), and HR-pyPopStar (HR - \citealp{millan2021hr}). By analysing the age and metallicity radial profiles from different stellar libraries, the model-based ones (SB99 and HR) do not present these young metal-poor stellar populations, since they have the proper grid to fit them properly. Furthermore, Boecker et al. (in preparation, private communication) show that these regions spatially match Wolf-Rayet stars' emission, which are not included in the empirical libraries but are in the model-based ones. Nevertheless, the general radial trends of MILES, XSL, and SB99 are similar, and the median values for MILES and SB99 in the remaining regions are similar as well. These lead us to believe that the choice of stellar library will not impact our general results and analysis regarding the sky subtraction (see Section~\ref{sec_results}). To understand how much these regions might affect our results, in this Appendix we repeat the analysis from Fig~\ref{fig_FLUXRatio_NGC1097} and Fig~\ref{fig_R_NGC1097} in Fig.~\ref{fig_FLUXRatio_NGC1097_MPmask}, but masking the regions in which the metallicity is below -0.4, highlighted in Fig.~\ref{fig_NGC1097_MPmask} -- the masked regions correspond to 15\% of the spaxels in the cube. It is clear that, qualitatively, our results stand, and these regions have very little influence on our main conclusions.

\section{Spectral analysis}
\label{appendix_b}

\begin{figure*}
        \centering
        \includegraphics[width=0.99\linewidth]{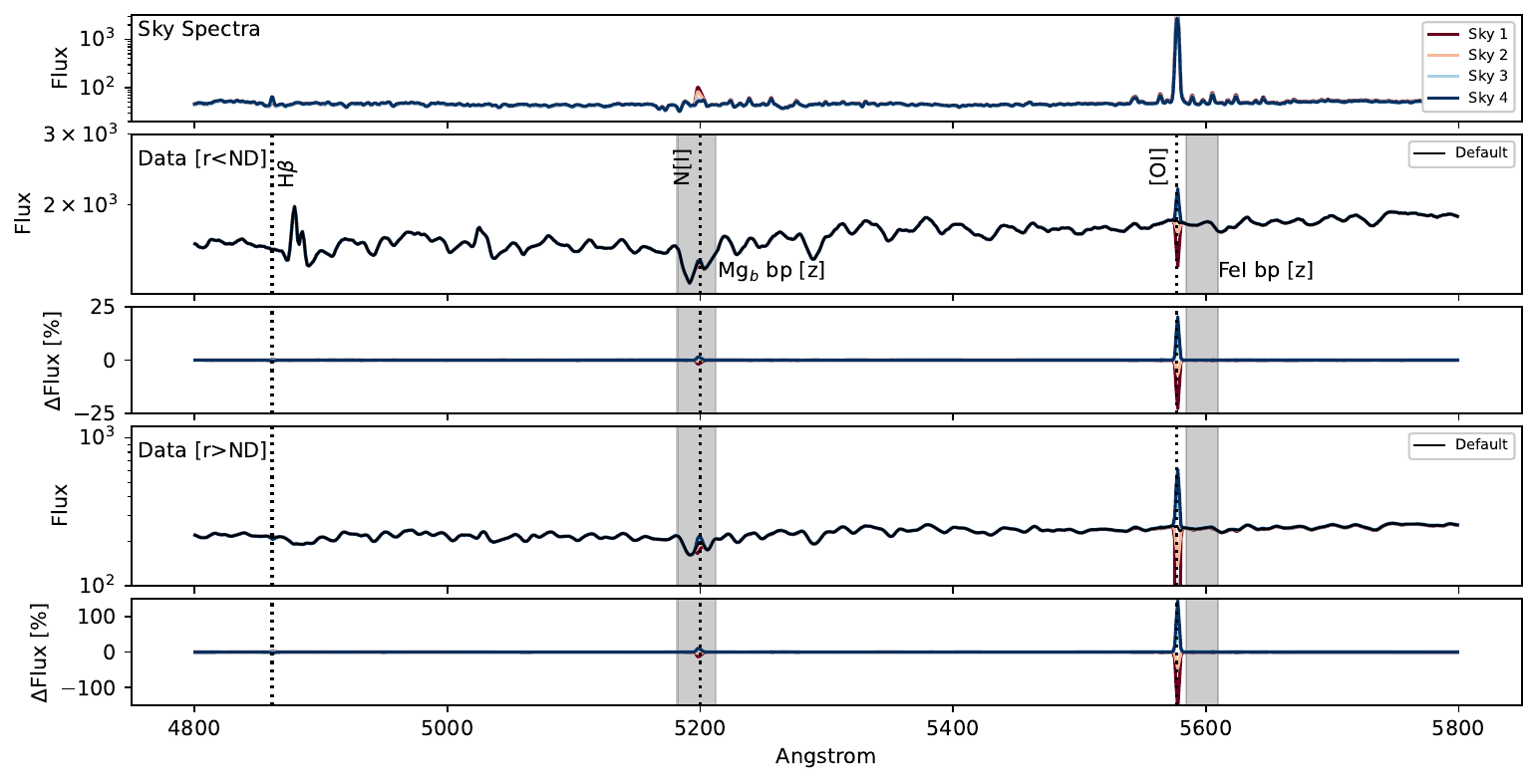}
        \caption{{Spectral analysis of NGC\,1097:} From top to bottom, we display the sky spectra, the reduced target spectra inside the nuclear disc, the difference with respect to the default reduction, the reduced target spectra outside the nuclear disc, and the corresponding difference with respect to the default reduction. Flux units are $10^{-20} \, \mathrm{erg \, s^{-1} \, cm^{-2} \, \mathring{A}^{-1}}$. We highlight the most heavily affected features, H$\beta$, [N\,{\sc i}], and [O\,{\sc i}], along with the Mg$_{b}$ and Fe\,{\sc i} index bandpasses (shaded grey areas), which are fundamental for measurements of metallicity and $\alpha$-enhancement -- zoomed-in spectra are available in Figure~\ref{fig_spectra_NGC1097_5800_zoom}.}
        \label{fig_spectra_NGC1097_5800}
\end{figure*}

In this letter, we considered a narrow region of the spectrum to perform this analysis ($4800\text{--}5800\,\text{\AA}$), as it is generally expected not to be heavily affected by sky emission lines (e.g., \citealp{bittner2020inside,gonccalves2020well}); however, we demonstrated that this is not the case in Section~\ref{sec_results}. To further investigate the origin of the differences found, in Figures~\ref{fig_spectra_NGC1097_5800} and \ref{fig_spectra_NGC1433_5800} we compare the sky spectra for the different sky frames (see Figure~\ref{fig_1}), the median target spectra inside and outside the nuclear disc, and the respective differences in relation to the default reduction for NGC\,1097 and NGC\,1433, respectively. It is clear that the continuum slope of the spectra is not strongly affected, but rather  \textcolor{black}{the continuum flux level} and/or specific lines: H$\beta$, [N\,{\sc i}], and [O\,{\sc i}], both inside and outside the nuclear disc, albeit with different intensities. Below, we discuss how these features impact the derived stellar population properties.

\begin{itemize}
    \item \textbf{H$\beta$:} Among the altered spectral features, the least affected is H$\beta$, which is classically associated with stellar population \textbf{age} indicators. Even though this line is only moderately affected in both galaxies, its relative weight is significantly different inside and outside the nuclear disc -- i.e., in bright and faint regions. For the region inside the nuclear disc of NGC\,1097, the flux difference is within $0.25\%$, while outside it reaches $2.5\%$, which is ten times higher. For NGC 1433, H$\beta$ does not seem to be affected inside the nuclear disc, whereas outside it reaches error values of up to $5\%$. This explains why the retrieved ages remain consistent for regions with SNR $>50$, but diverge for fainter regions
    
    \item \textbf{[O\,{\sc i}]:} The line most heavily affected in this exercise is the forbidden [O\,{\sc i}] emission line, which is strongly associated with atmospheric airglow. For NGC\,1097, the over-subtraction of [O\,{\sc i}] produces an artificial absorption feature in skies 1 and 2 (which possess stronger sky emission), while yielding an emission feature for skies 3 and 4. The corresponding flux difference is within $25\%$ inside the nuclear disc and up to $100\%$ outside it. Meanwhile, for NGC\,1433, the differences are within $25\%$ outside the nuclear disc when compared to the default reduction. This can result in a systematic effect on \textbf{metallicity} measurements, since [O\,{\sc i}] falls directly next to the redshifted Fe\,{\sc i}$\lambda5573$ index bandpass. It can affect the local pseudo-continuum bandpasses. 
    
    \item \textbf{[N\,{\sc i}]}: Similar to [O\,{\sc i}], the [N\,{\sc i}] line is located within the Mg$_{b}$ redshifted bandpass, which can affect both the measurements of \textbf{metallicity} and \textbf{$\alpha$-enhancement}. A word of caution: this is the case for galaxies with redshifts between 0.0014 and 0.0077. As expected, the flux difference for both NGC\,1097 and NGC\,1433 is within $1\%$ inside the nuclear disc, but reaches up to $10\%$ outside it.
\end{itemize}

Lastly, for NGC\,1433, it is clear that the continuum level changes between the nights: around $3\%$ within the nuclear disc, and between $1-7\%$ outside the nuclear disc. By changing the level of the continuum, the strength of absorption lines is also affected, resulting in the different stellar populations derived for each night. 

While we restricted our analysis to a small part of the spectrum, many studies consider a wider range, especially towards redder wavelengths. It is expected that the sky affects the continuum slope of the spectrum in redder regions more strongly, potentially producing more extreme results for the retrieved stellar ages. As explicitly stated in Section~\ref{sec_methodology}, an extensive analysis of this is beyond the scope of this letter. However, we performed a quick visual analysis of the entire spectra of both galaxies to identify whether or not their slopes are strongly affected. Evidently, there are more emission lines associated with the sky at $\lambda > 5800\,\text{\AA}$, which can potentially affect the results even further if they are not properly accounted for. However, the differences in flux, both inside and outside the nuclear disc, do not present a noticeable slope variation that could further impact the stellar population age analysis.

\begin{figure*}
        \centering
        \includegraphics[width=0.99\linewidth]{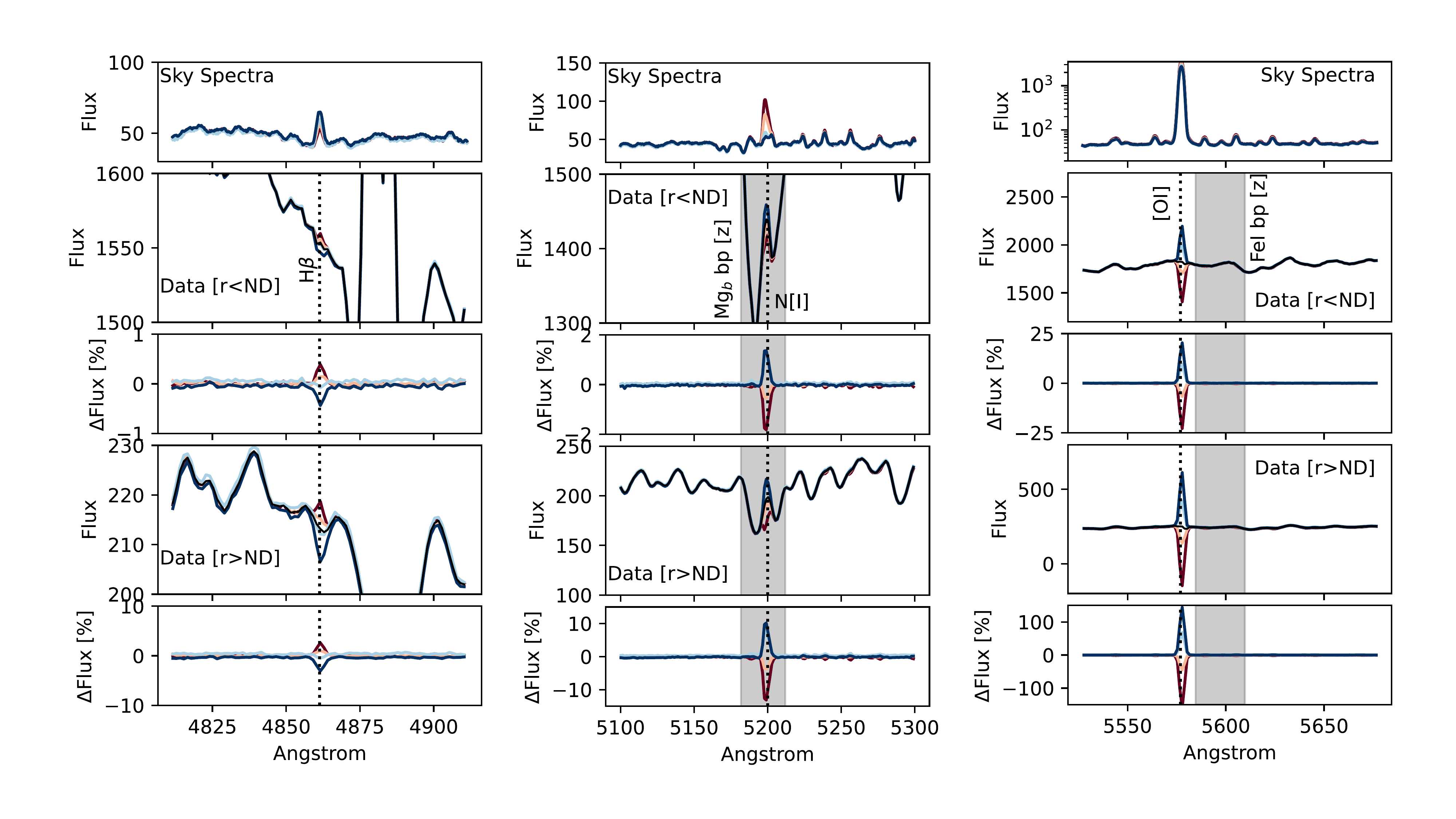}
        \caption{Same as Figure~\ref{fig_spectra_NGC1097_5800}, zooming in around regions of interest.}
        \label{fig_spectra_NGC1097_5800_zoom}
\end{figure*}

\begin{figure*}
        \centering
        \includegraphics[width=0.99\linewidth]{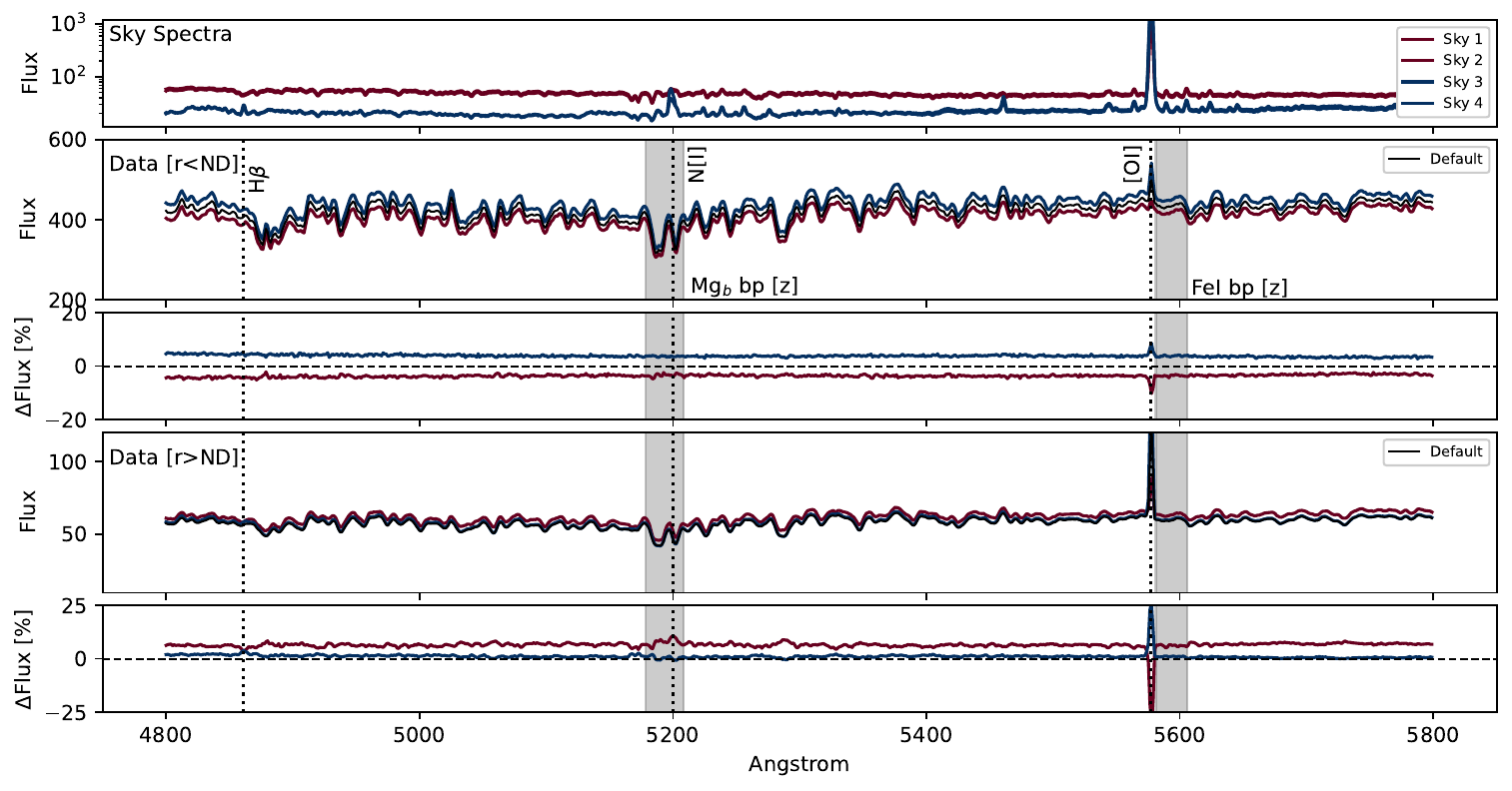}
        \caption{{Spectral analysis of NGC\,1433:} Similar to Figure~\ref{fig_spectra_NGC1097_5800}, but for NGC\,1433. For the region inside the nuclear disc, H$\beta$, Mg$_{b}$, and [N\,{\sc i}] do not seem to be affected, with only [O\,{\sc i}] being mildly impacted. Outside the nuclear disc, however, these features are not only heavily affected but also appear to invert, which can systematically alter the measurements of age, metallicity, and $\alpha$-enhancement witnessed in Figure~\ref{fig_R_NGC1433} -- zoomed-in spectra are available in Figure~\ref{fig_spectra_NGC1433_5800_zoom}}
        \label{fig_spectra_NGC1433_5800}
\end{figure*}

\begin{figure*}
        \centering
        \includegraphics[width=0.99\linewidth]{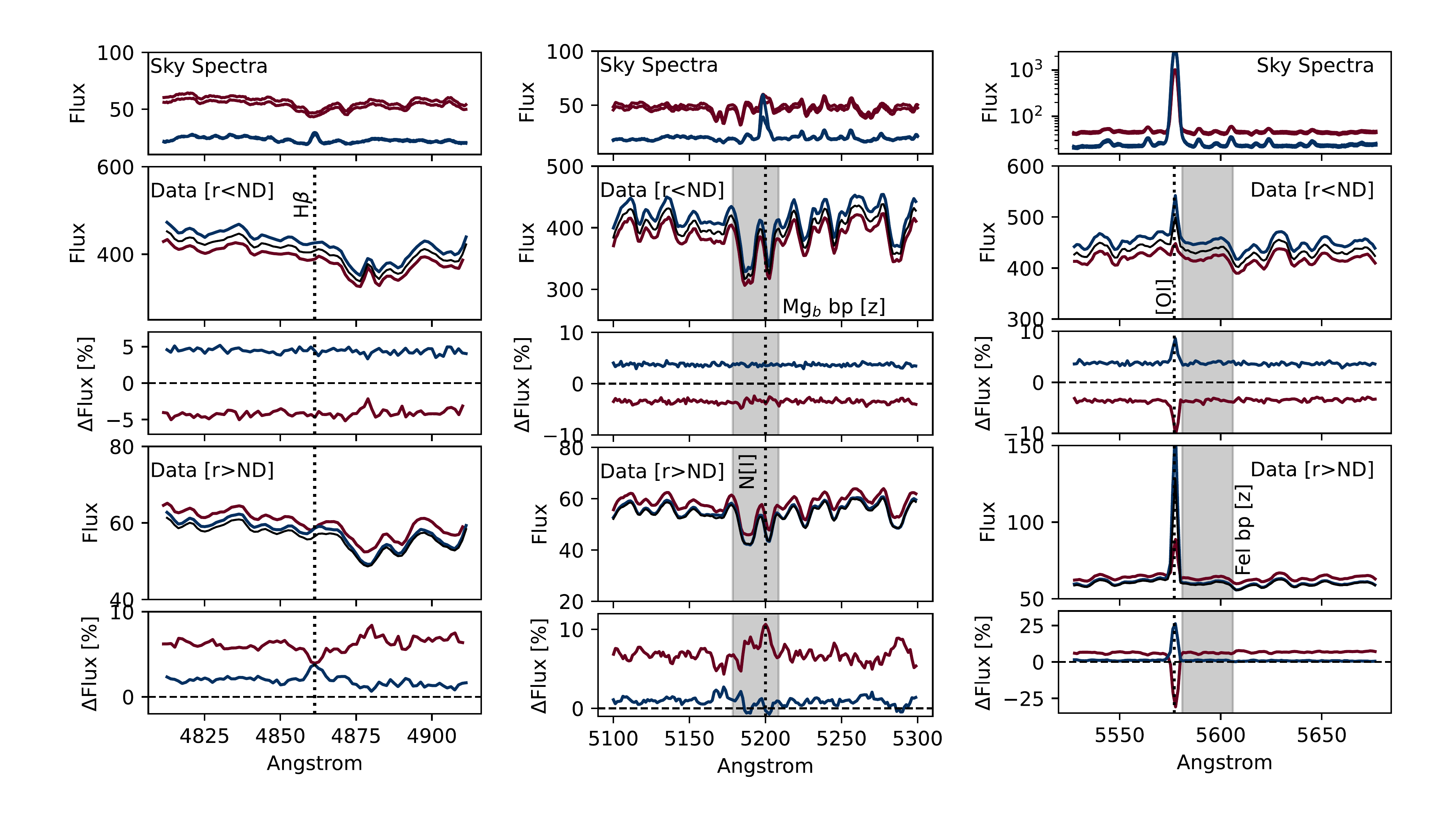}
        \caption{Same as Figure~\ref{fig_spectra_NGC1433_5800}, zooming in around regions of interest.}
        \label{fig_spectra_NGC1433_5800_zoom}
\end{figure*}
\end{appendix}

\end{document}